\documentclass[aps,twocolumn,showpacs]{revtex4}
\usepackage{graphicx}
\usepackage{hyperref}
\usepackage{color}
\usepackage{float}
\begin{document}

\title{Detection of the interfacial exchange field at a ferromagnetic insulator-nonmagnetic metal interface with pure spin currents.}

\author{P. K. Muduli$^{1}$}\email{muduli.ps@gmail.com}
\author{ M. Kimata$^{1}$}\altaffiliation[Present address: ]{Institute for Materials Research, Tohoku University, Sendai 980-8577, Japan.}
\author{ Y. Omori$^{1}$}
\author{T. Wakamura$^{1}$}\altaffiliation[Present address: ]{Laboratoire de Physique des Solides, Universite Paris-Sud, 91400, Orsay, France.}
\author{Saroj P. Dash$^3$}
\author{YoshiChika Otani$^{1,2}$}\email{yotani@issp.u-tokyo.ac.jp} \affiliation{ $^1$Institute for Solid State Physics, University
of Tokyo, Kashiwa 277-8581, Japan} \affiliation{$^2$Center for
Emergent Matter Science, RIKEN, 2-1 Hirosawa, Wako 351-0198,
Japan} \affiliation{ $^3$Department of Microtechnology and
Nanoscience, Chalmers University of Technology, SE-41296 Goteborg,
Sweden}

\date{\today}
\begin{abstract}
At the interface between a nonmagnetic metal (NM) and a
ferromagnetic insulator (FI) spin current can interact with the
magnetization, leading to a modulation of the spin current. The
interfacial exchange field at these FI-NM interfaces can be probed
by placing the interface in contact with the spin transport
channel of a lateral spin valve (LSV) device and observing
additional spin relaxation processes. We study interfacial
exchange field in lateral spin valve devices where Cu spin
transport channel is in proximity with ferromagnetic insulator EuS
(EuS-LSV) and yttrium iron garnet Y$_3$Fe$_5$O$_{12}$ (YIG-LSV).
The spin signals were compared with reference lateral spin valve
devices fabricated on nonmagnetic Si/SiO$_2$ substrate with MgO or
AlO$_x$ capping. The nonlocal spin valve signal is about 4 and 6
times lower in the EuS-LSV and YIG-LSV, respectively. The
suppression in the spin signal has been attributed to enhanced
surface spin-flip probability at the Cu-EuS (or Cu-YIG) interface
due to interfacial spin-orbit field. Besides spin signal
suppression we also found widely observed low temperature peak in
the spin signal at $T \sim$30 K is shifted to higher temperature
in the case of devices in contact with EuS or YIG. Temperature
dependence of spin signal for different injector-detector
distances reveal fluctuating exchange field at these interfaces
cause additional spin decoherence which limit spin relaxation time
in addition to conventional sources of spin relaxation. Our
results show that temperature dependent measurement with pure spin
current can be used to probe interfacial exchange field at the
ferromagnetic insulator-nonmagnetic metal interface.

\end{abstract}
\maketitle
\clearpage
\section{Introduction}

A strong Rashba-type interfacial spin-orbit field is a common
feature at the interface between different materials due to the
inversion symmetry breaking. Following the discovery of
spin-to-charge conversion at the interface via the inverse
Rashba-Edelstein effect (IREE), currently there is a great deal of
interest to study different kind of metal-insulator interface with
Rashba spin-orbit
interaction\cite{otani,Soumyanarayanan,Hellman,edelstein,rojasan}.
In particular, an understanding of the interfacial spin-orbit and
exchange field at the interface between a nonmagnetic metal (NM)
and a ferromagnetic insulator (FI) is of crucial importance for
many spintronic phenomena such as spin pumping\cite{Tserkovnyak},
spin-transfer torque\cite{Ralph}, spin Seebeck
effect\cite{Uchida}, spin Peltier effect\cite{Flipse}, spin Hall
magnetoresistance (SMR)\cite{Nakayama}, and magnon-mediated
magnetoresistance (MMR)\cite{Goennenwein}. These effects rely on
the transfer of spin angular momentum from the conduction-electron
spins in the nonmagnetic metal to the magnetization of
ferromagnetic insulator at the FI/NM interface. Therefore, the
relative orientation between the magnetization of the FI and the
spin polarization of spin current dictates the interfacial spin
transmission. In conventional spin pumping theory these
interfacial spin transfer effects is usually described by a
complex quantity called spin mixing conductance $G_{ \uparrow
\downarrow } ( = G_r + iG_i )$\cite{Tserkovnyak}. The effect of
interfacial exchange field is a priori included in the imaginary
part  of spin mixing conductance $G_i$. However, $G_i$ is usually
ignored as it is an order of magnitude smaller compared to the
real part $G_r$. In spite of pronounced progress in these studies
several aspects of the spin-transfer mechanism at the FI/NM
interface remains unanswered. So far, most of the experiments on
spin-angular-momentum transfer via the interfacial exchange
interaction have been focused on the spin transfer to adjacent
nonmagnetic heavy metals with strong spin-orbit coupling (SOC).
Here, the SOC transforms the spin current into a measurable
electric voltage via the inverse spin Hall effect (ISHE). However,
in this kind of studies the influence of the interface cannot be
disentangled from spurious bulk effects resulting from
sample-to-sample variation of spin Hall angle and spin diffusion
length of the heavy metal. Interfacial exchange field can be
studied more appropriately in a FI/NM combination with weak
spin-orbit NM like Cu, Ag and Al. Due to large spin diffusion
length of these metals, the conduction electrons traveling near
the FI/NM interface experience exchange interaction from the
localized magnetic moments of the ferromagnetic insulator and
become spin decoherent. This spin information loss at the FI/NM
interface caused by the interfacial spin-orbit and exchange fields
can be measured  with a lateral spin valve device.

Lateral spin valve (LSV) is a rudimentary spintronics device that
offer straight forward method to study spin transport due to it's
unique ability to separate charge and spin currents. It is one of
the most unambiguous techniques for probing the spin transport in
nonmagnetic materials with weak spin-orbit coupling. In lateral
spin valve experiments non-equilibrium spin distribution is
generated inside a nonmagnetic material by passing current through
a ferromagnet. This spin distribution is transported between two
ferromagnetic electrodes in a material with weak SOC, such as Cu,
Ag or Al. In nonmagnetic metals the spin relaxation of conduction
electrons is usually described by Elliott-Yafet (EY)
mechanism\cite{Elliott,Yafet}. According to this theory, spin
relaxes by momentum scattering events from phonons, impurities,
boundaries and interfaces. These relaxation processes can be
quantified using the spin-flip time $\tau_{sf}$. At high
temperature the spin relaxation is predominantly caused by
phonons, while at low temperatures spin-flip processes due to the
impurity and surface scattering dominate. Therefore, spin
relaxation time measurement at low temperature in LSVs provide
unique opportunity to study interfacial spin-orbit and exchange
fields. Although these fields are readily resolved in graphene and
other 2D materials, in thick metallic nanowires they have been
ignored so far. Recently magnetically controlled modulation of the
spin current was observed in Cu and Al spin transport channel in
contact with ferrimagnetic yttrium iron garnet Y$_3$Fe$_5$O$_{12}$
(YIG) substrate\cite{Villamor-yig,Dejene}. This finding has
renewed the search for an effective way to regulate spin current
in metallic nanowires in contact with ferromagnetic insulators.

So far most of the study on FI-NM interface has been conducted on
YIG-NM interfaces, where the $s-d$ exchange interactions dominate
at the interface\cite{Maekawa-rev}. Better understanding of FI-NM
interface can emerge if similar study is done with fundamentally
different kind of FI-NM interface. The magnetism in EuS is
determined by indirect $s-f$ exchange interactions\cite{Miao-rev},
which makes the EuS-NM interface fundamentally different to YIG-NM
interface. Experimentally, EuS is among the most popular magnetic
insulators which has shown to provide exceptional value of
interfacial exchange field. Large magnetic moment of Eu$^{2+}$
($S_z$ $\sim$ 7$\mu_B$) along with large exchange coupling
constant ($J$ = 10 meV) leads to a huge interfacial exchange field
($E_{ex}$$\propto$ $JS_z$). Magnetic proximity effect of EuS on
superconductors has been demonstrated experimentally by many
groups\cite{Strambini,moodera,hao}. Proximity-induced
ferromagnetism in graphene and topological insulators in contact
with EuS has also been reported\cite{pwei,wei,zhao,Katmis}. More
recently, considerably large interfacial exchange field up to 14 T
has been estimated to exist at graphene-EuS interface\cite{wei}.
In contrast, a smaller exchange field of the order of $\sim$0.2 T
has been observed experimentally in the graphene-YIG
interface\cite{Leutenantsmeyer}. Therefore EuS-NM interface
provides a good system to study interfacial exchange field.
Furthermore, EuS thin films can be grown ultra-thin down to 1 nm
and still retain magnetic property. Which makes it a very
promising substitute for YIG in nanoscale devices.

In this paper, we systematically investigated the spin transport
properties of lateral spin valve devices in which spin transport
channel is in contact with ferromagnetic insulators EuS and YIG.
Spin current was created in a Cu bridge between two
Ni$_{80}$Fe$_{20}$ (Py) electrodes by non-local method.
Temperature evolution of spin signal of these LSV devices were
examined in order to distinguish effects arising from interfacial
spin-orbit and exchange fields. This spin signal was compared with
identical measurements on reference lateral spin valves in contact
with nonmagnetic insulators like MgO and AlO$_x$. We provide
plausible explanation for anomalous temperature evolution of spin
signal in the EuS-LSV and YIG-LSV devices incorporating
fluctuating exchange field at the ferromagnetic
insulator-nonmagnetic metal interface which can cause additional
temperature-dependent spin relaxation.

\section{Experimental details}

Ultrathin EuS films were deposited by e-beam evaporation with slow
growth rate $\sim$0.13 ${\AA}$/s under a base pressure of $\sim$ 2
$\times$ 10$^{-8}$ torr. Magnetic property of EuS thin films in
the thickness range 2-10 nm were characterized with Quantum Design
MPMS magnetometer. Lateral spin valves were fabricated by
multi-step e-beam lithography and lift-off process.  First 30 nm
thick two Permalloy (Ni$_{80}$Fe$_{20}$) electrodes were deposited
followed by 3 nm MgO in-situ by e-beam evaporation. Then 100 nm
thick Cu spin transport channel was deposited in another UHV
chamber by a Joule heating evaporator using 99.9999 $\%$ purity Cu
source. Prior to Cu deposition samples were Ar$^+$-ion milled for
40 sec to remove resist left-over.  After Cu deposition devices
were transferred to another UHV chamber and 5 nm EuS capping layer
was deposited following 40 sec in-situ Ar$^+$-ion milling. After
Cu lift-off the EuS capped samples were further capped with 2 nm
AlO$_x$ by rf-sputtering to protect it against oxidation. EuS
capped sample without subsequent AlO$_x$ capping showed rapid
degradation with time. In all other spin valve devices AlO$_x$ or
MgO capping was done ex-situ after Cu lift-off without any
Ar$^+$-ion cleaning process. Therefore presence of native oxide
layer on Cu nanowire cannot be ruled out. Another series of
lateral spin valves were also fabricated on Si/SiO$_2$/EuS(10 nm)
thin film (labeled as EuS substrate) and 2 micrometer YIG film
grown on gadolinium gallium garnet (GGG) substrate (labeled as YIG
substrate) following similar procedure and capped with AlO$_x$ at
the end. The YIG  thin film grown on GGG substrate was obtained
from Innovent e. V. (Jena, Germany). A typical lateral spin valve
device with nonlocal measurement configuration is shown in Fig.
1(b). To have different switching fields in two Py nanowires ends
of the first Py wire are connected to two large pads which enable
the nucleation of domain wall at lower magnetic fields than the
thin long Py nanowire. The nonlocal measurements were performed in
a continuous flow cryostat using a phase sensitive lock-in
technique (modulation frequency $f$ = 173 Hz). All the temperature
dependent measurements were done in a single heating cycle from 2
to 300 K. For accurate calculation of resistivity dimensions of Cu
and Py electrodes were measured by Scanning Electron Microscope
(SEM) for each device.

\begin{widetext}
\begin{center}
\begin{figure}[H]
\begin{tabular}{ll}
  \centering
  \includegraphics[width= 7 cm]{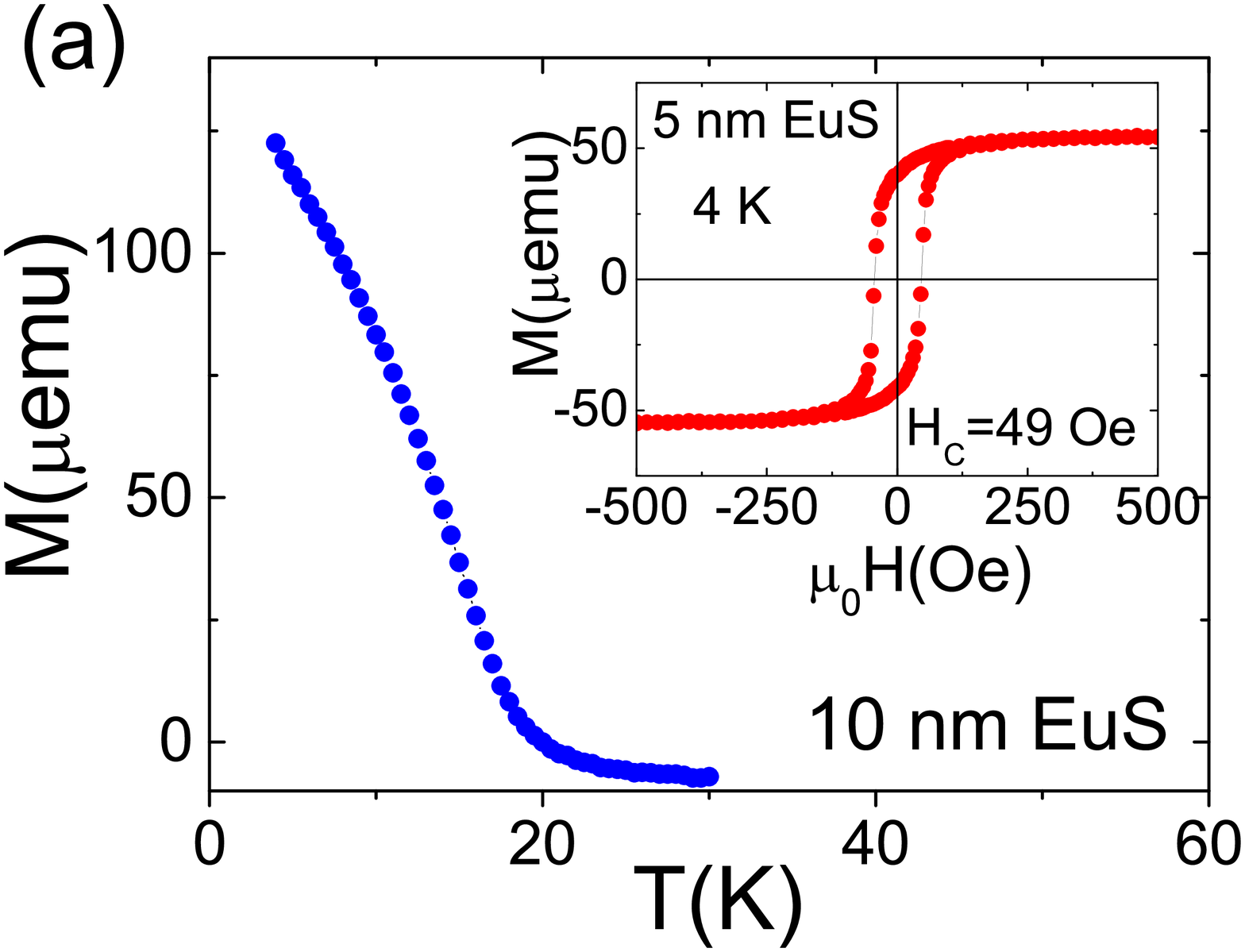}
&

  \includegraphics[width= 6.5 cm]{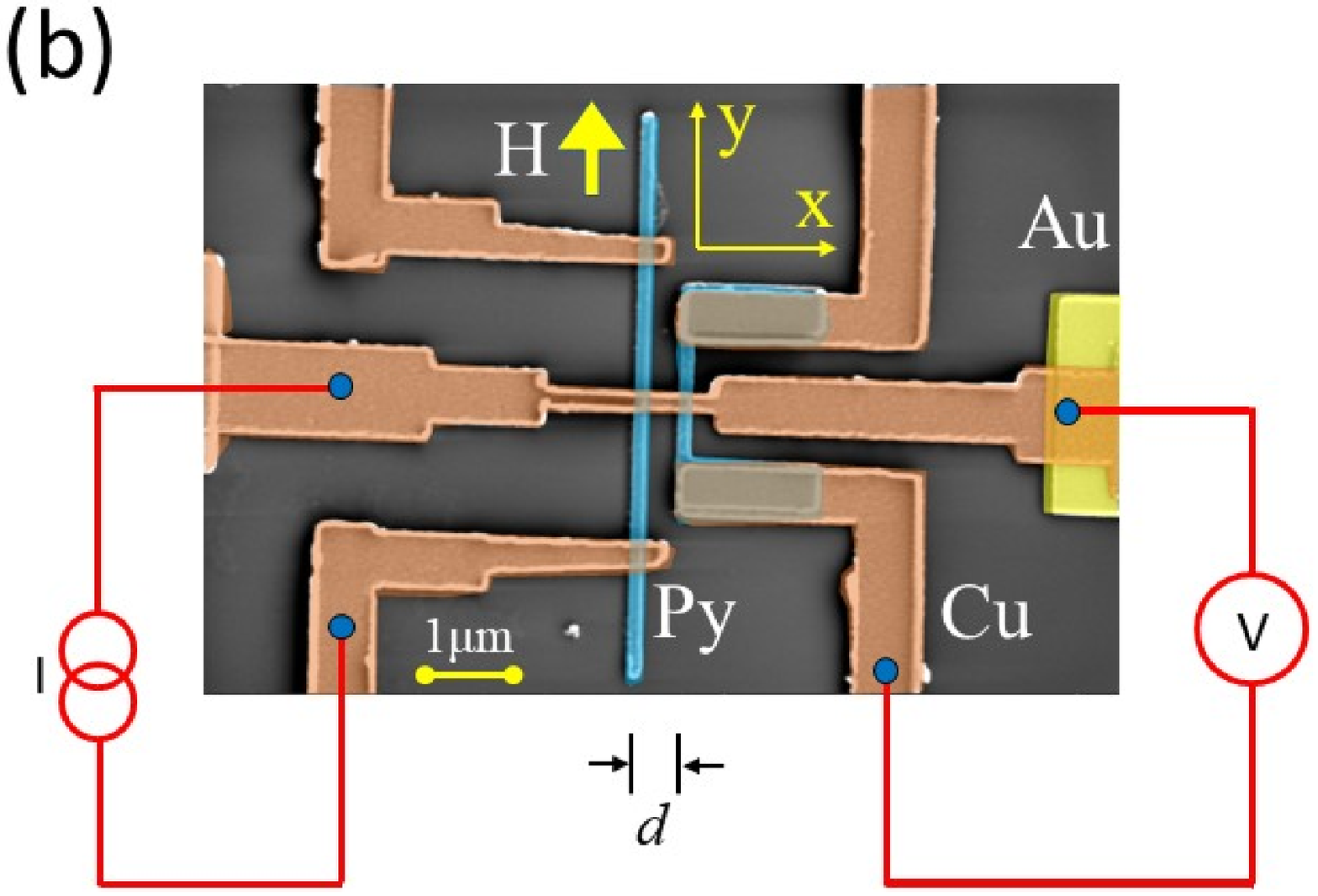}\\

 \centering
 \includegraphics[width= 7 cm]{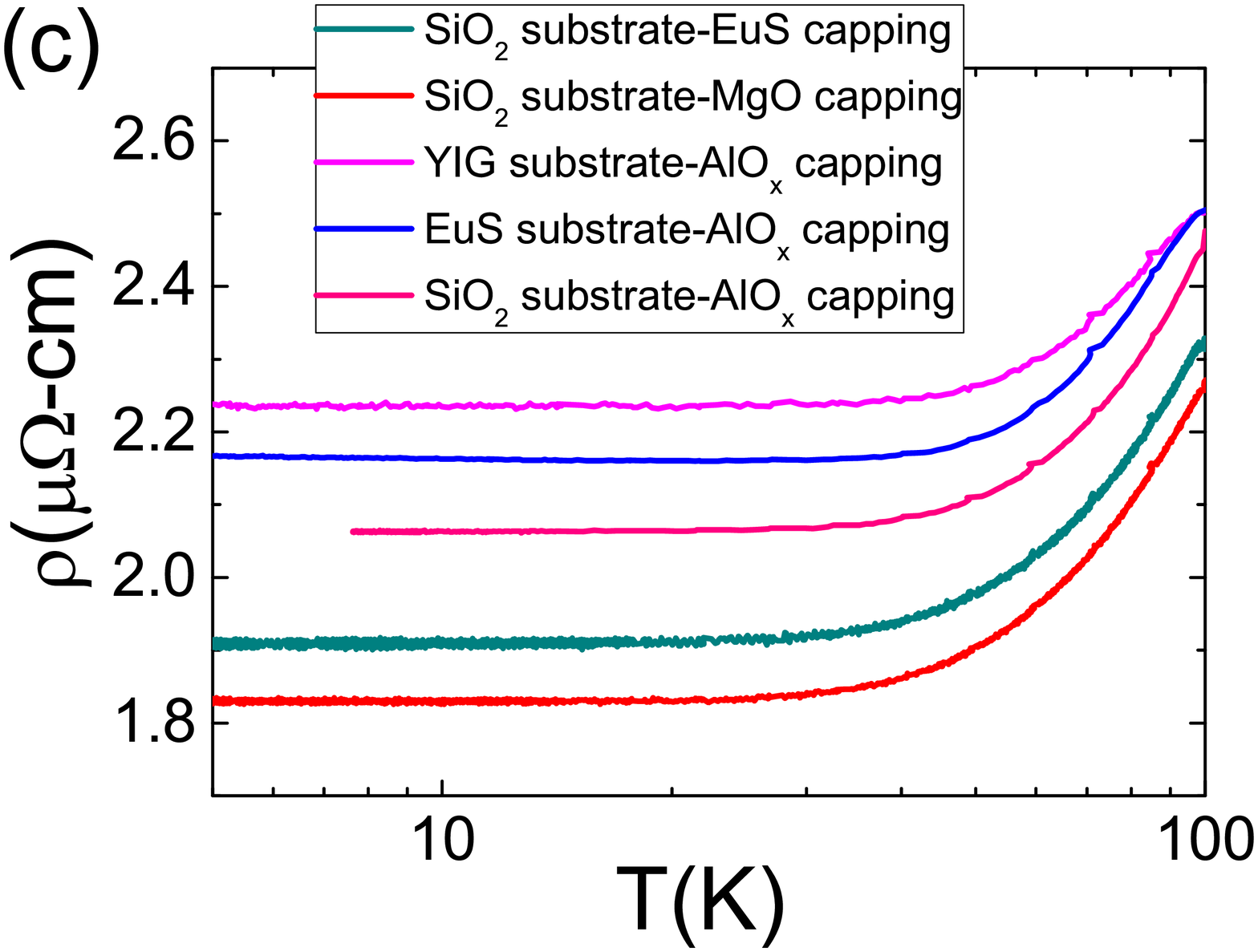}
&

 \includegraphics[width= 7 cm]{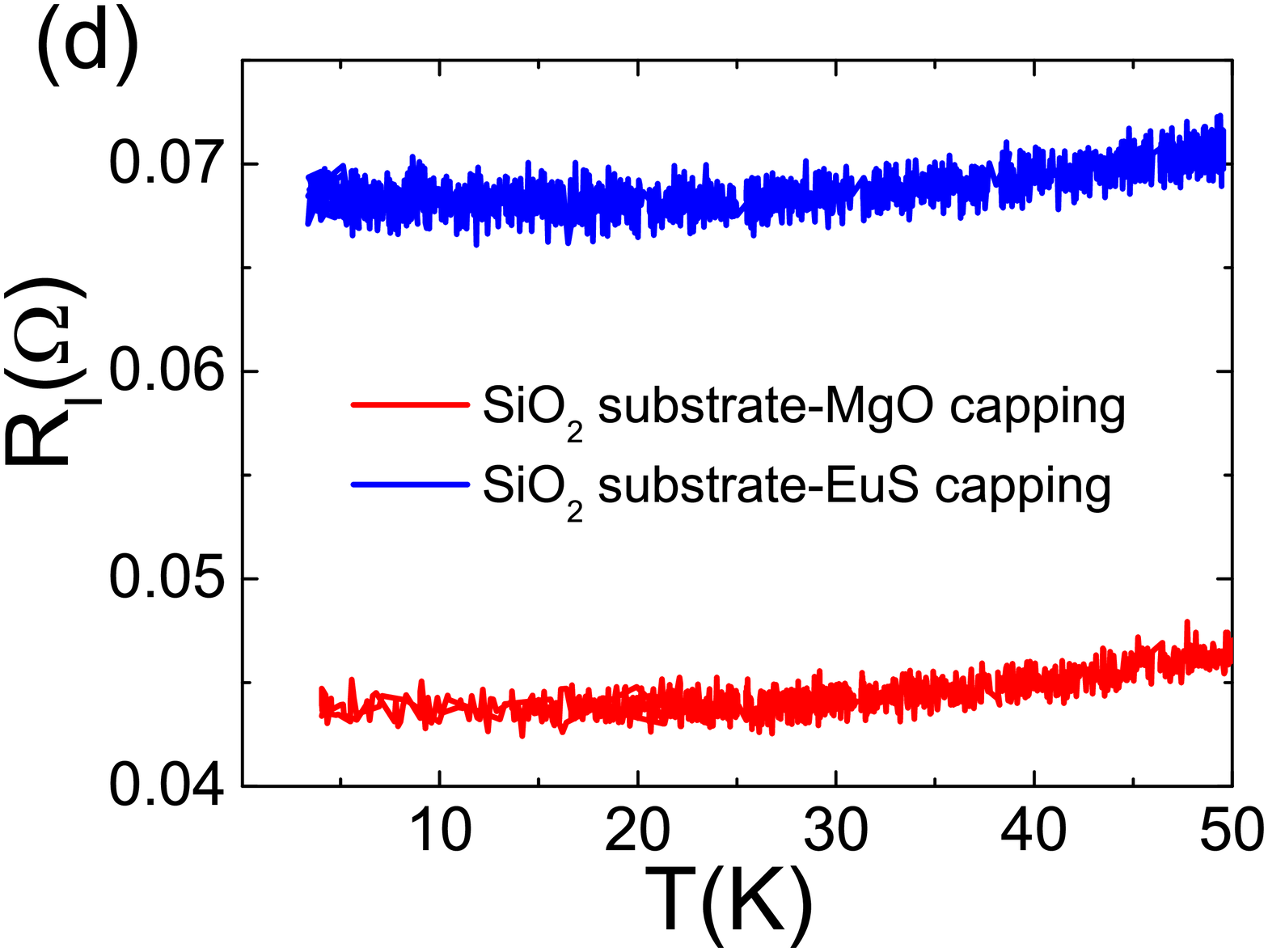}\\

\end{tabular}
\caption{(a) Temperature-dependent magnetization of a 10 nm thick
EuS film. The inset shows M-H loop of a $\sim$5 nm thick EuS film
measured at 4 K and the field applied in the film plane. (b) SEM
picture of the lateral spin valve device with nonlocal measurement
configuration. (c) Temperature dependence of resistivity in the
temperature range 5-100 K for the 100 nm thick Cu nanowire on
different substrates and capping. The absolute value of the
resistivity can have maximum error up to $\sim$17 $\%$ due to the
uncertainty associated with the estimation of dimension of the Cu
nanowire. Neither curve shows an upturn of the resistivity at low
temperatures (note the logarithmic temperature scale), and
therefore the Kondo effect plays a negligible role. (d)
Temperature dependence of Py-MgO-Cu interface resistance in the
temperature range 5-50 K. Interface resistance was measured in
four probe configuration. Again a resistance minimum associated
with Kondo effect is absent.}
\end{figure}
\end{center}
\end{widetext}

\section{Results}

Fig. 1(a) shows temperature dependence of magnetization of a 10 nm
thick EuS film deposited by e-beam evaporation. Bulk EuS is known
to be an ideal Heisenberg ferromagnetic semiconductor with Curie
temperature ($T_C$) of about 16.6 K and band gap of $\sim$1.65 eV
at room temperature\cite{zinn}. A broad ferromagnetic transition
with Curie temperature T$_C$ $\sim$15 K can be seen in Fig. 1(a).
This is comparable to values reported for EuS thin films by
different groups\cite{moodera,Idzuchi-EuS,wolf}. Inset shows
magnetization (M-H) loop of a 5 nm thick EuS film measured at 4 K
with the magnetic field applied in the film plane. A small
coercive field $\sim$49 Oe with an almost rectangular loop was
found supporting the ferromagnetic properties of the EuS film. We
found EuS thin films are magnetic even down to $\sim$2 nm. Films
were also found to be very smooth with rms roughness $\sim$0.54 nm
over an area of 1$\mu$m $\times$ 1$\mu$m. The resistivity of EuS
films were found to be $\rho_{EuS}$ $\sim$0.5 $\Omega$-cm at room
temperature. Fig. 1(c) shows temperature dependence of resistivity
of Cu spin transport channel in the temperature range 4-200 K with
different substrate and capping. The resistivity of Cu nanowire
was measured using a 4-point configuration in which current is
sent through the Py electrodes and voltage is measured along Cu
channel. Temperature axis is plotted in log-scale to identify
resistivity minimum considered as an evidence of Kondo effect. No
resistivity minimum was observed indicating negligible role of
Kondo effect and magnetic impurities. Temperature independent
residual resistivity suggest electrical transport in these Cu
nanowires arise primarily from scattering with defects, grain
boundaries and surface. In these lateral spin valve devices 3 nm
MgO layer was inserted between Py and Cu to increase spin
injection efficiency. This rules out possibility of ferromagnetic
species diffusing into the Cu nanowire which creates magnetic
scattering sites. In addition, Py and Cu were deposited in two
separate UHV chambers which limit possible interdiffusion of
magnetic elements at the interface. Fig. 1(d) shows temperature
dependence of Cu/MgO/Py interface resistance for LSVs with MgO and
EuS capping. No resistivity upturn was observed suggesting limited
interfacial interdiffusion. We found LSV devices fabricated
following similar process with no MgO tunnel barrier also showed
similar temperature dependence as in Fig. 1(b) (see Supplementary
SFig. 1\cite{supp}).

\begin{widetext}
\begin{center}
\begin{figure}[H]
\begin{tabular}{ll}
  \centering
  \includegraphics[width= 7 cm]{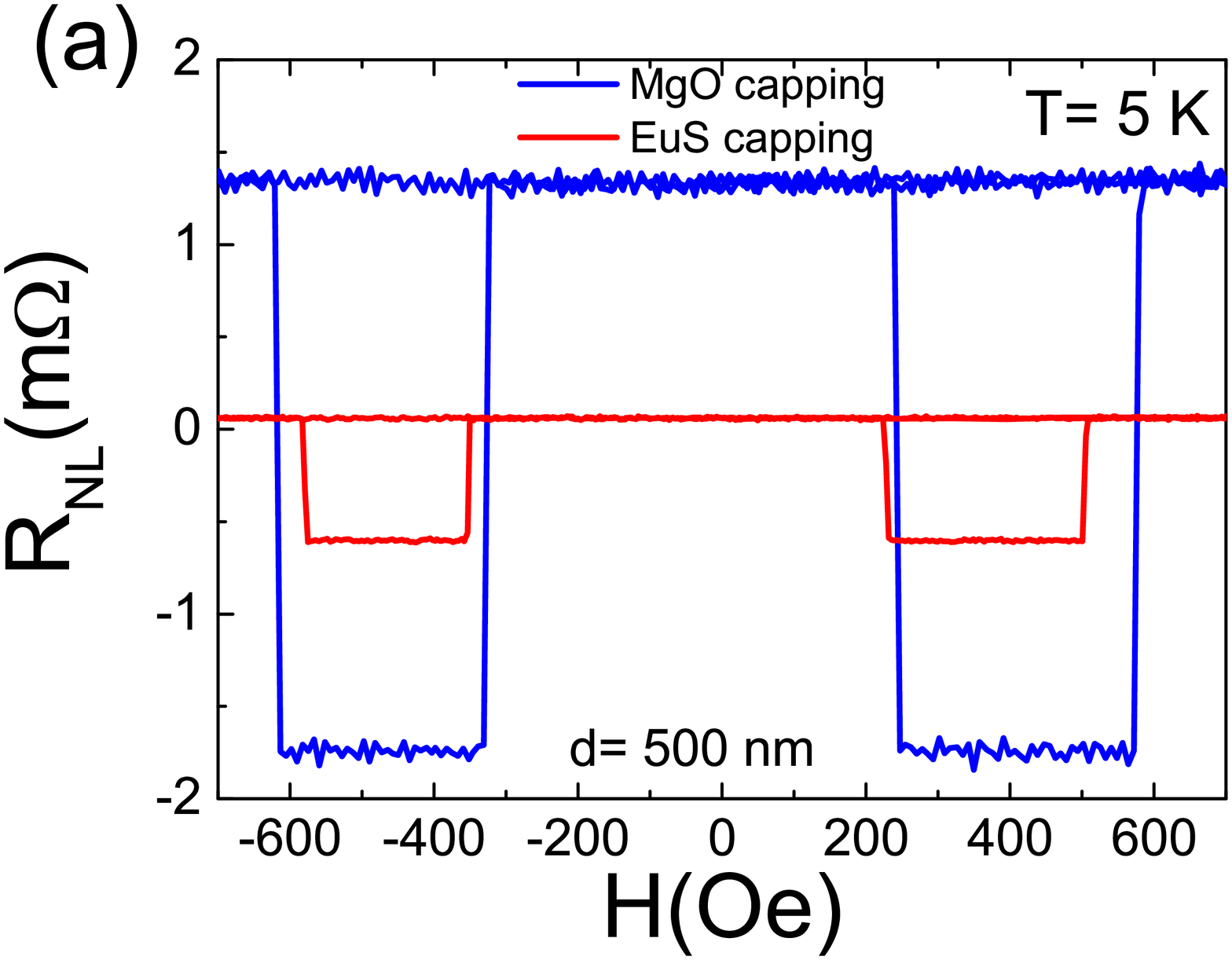}
&

  \includegraphics[width= 7 cm]{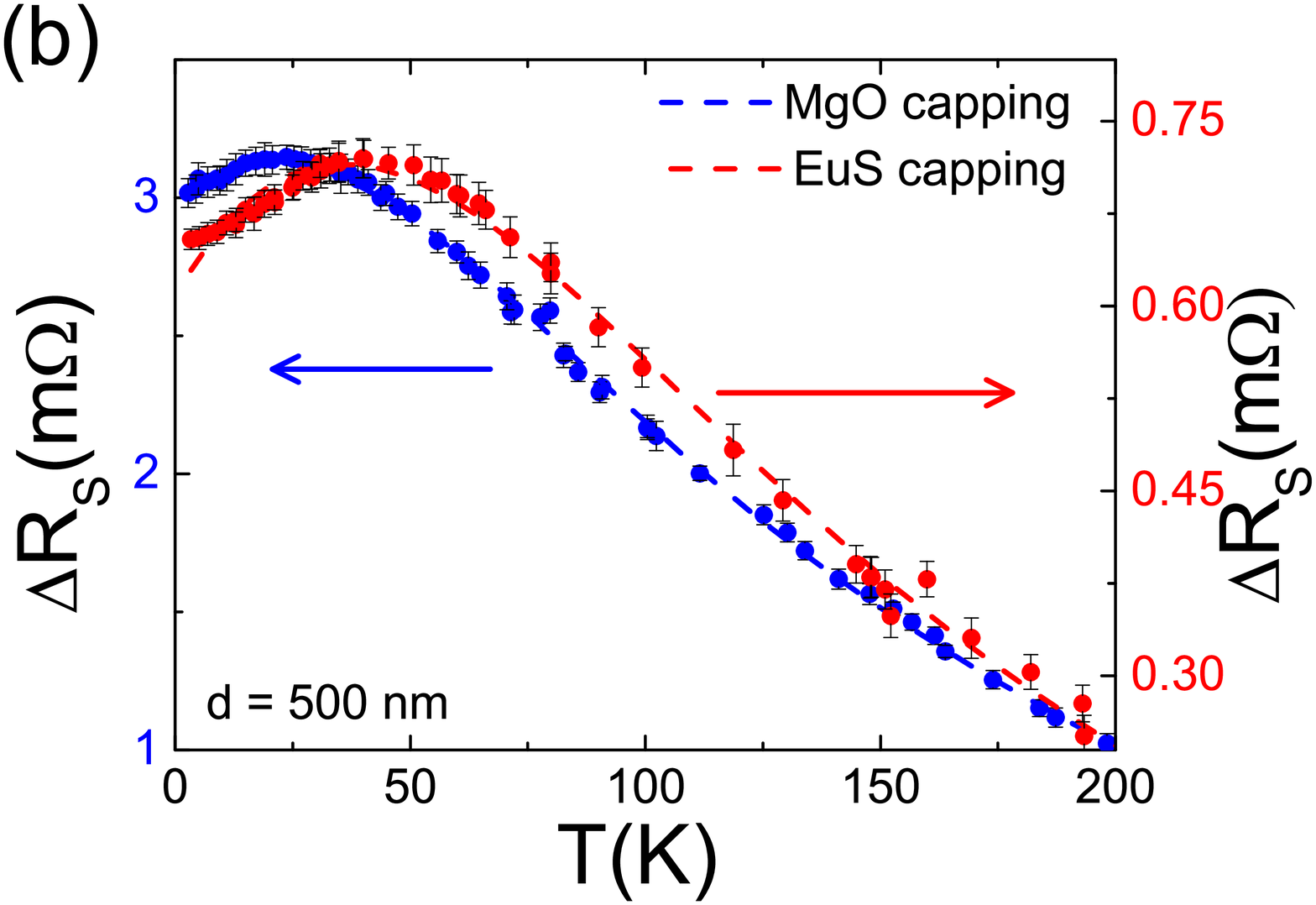}\\

 \centering
 \includegraphics[width= 7 cm]{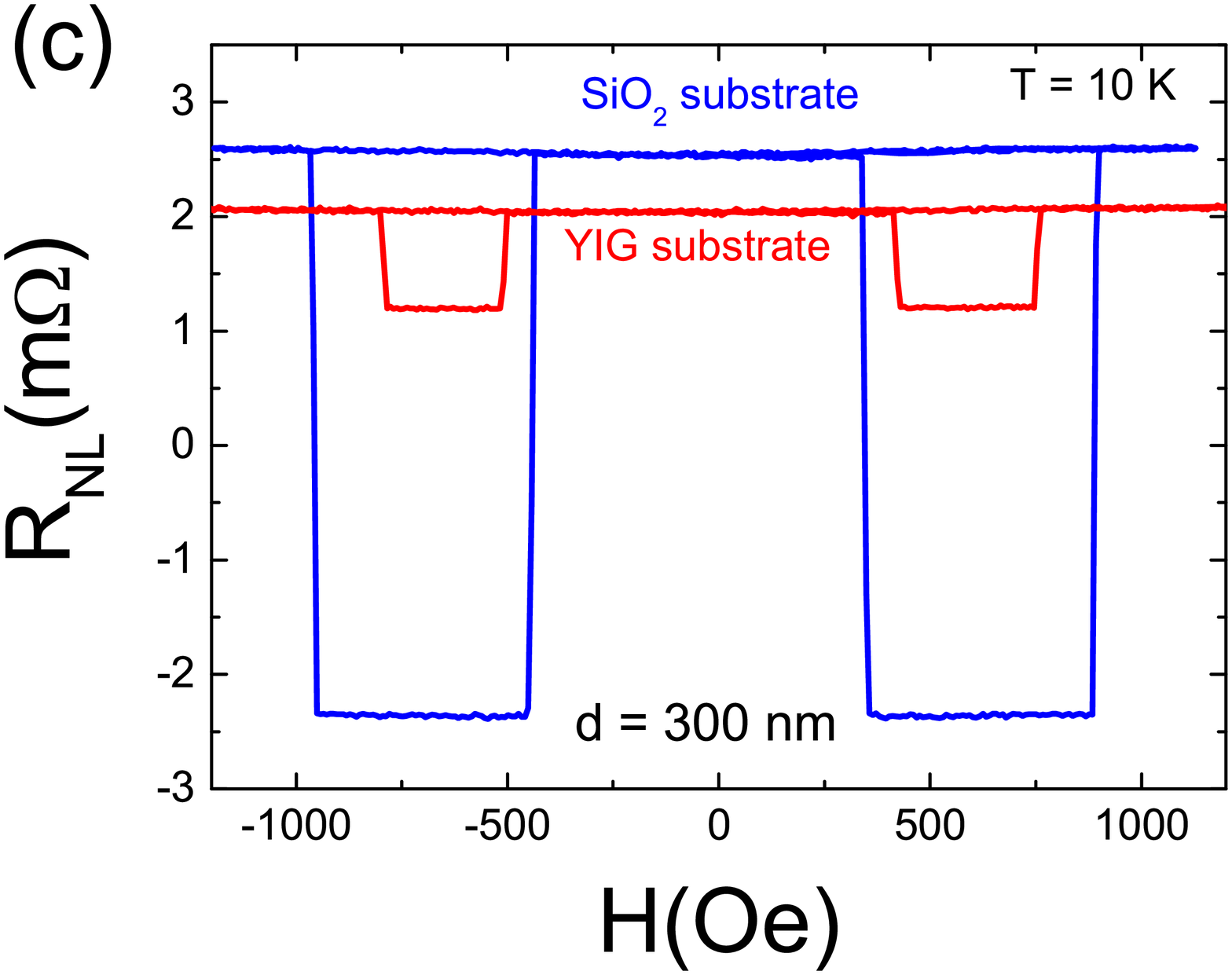}
&

 \includegraphics[width= 7 cm]{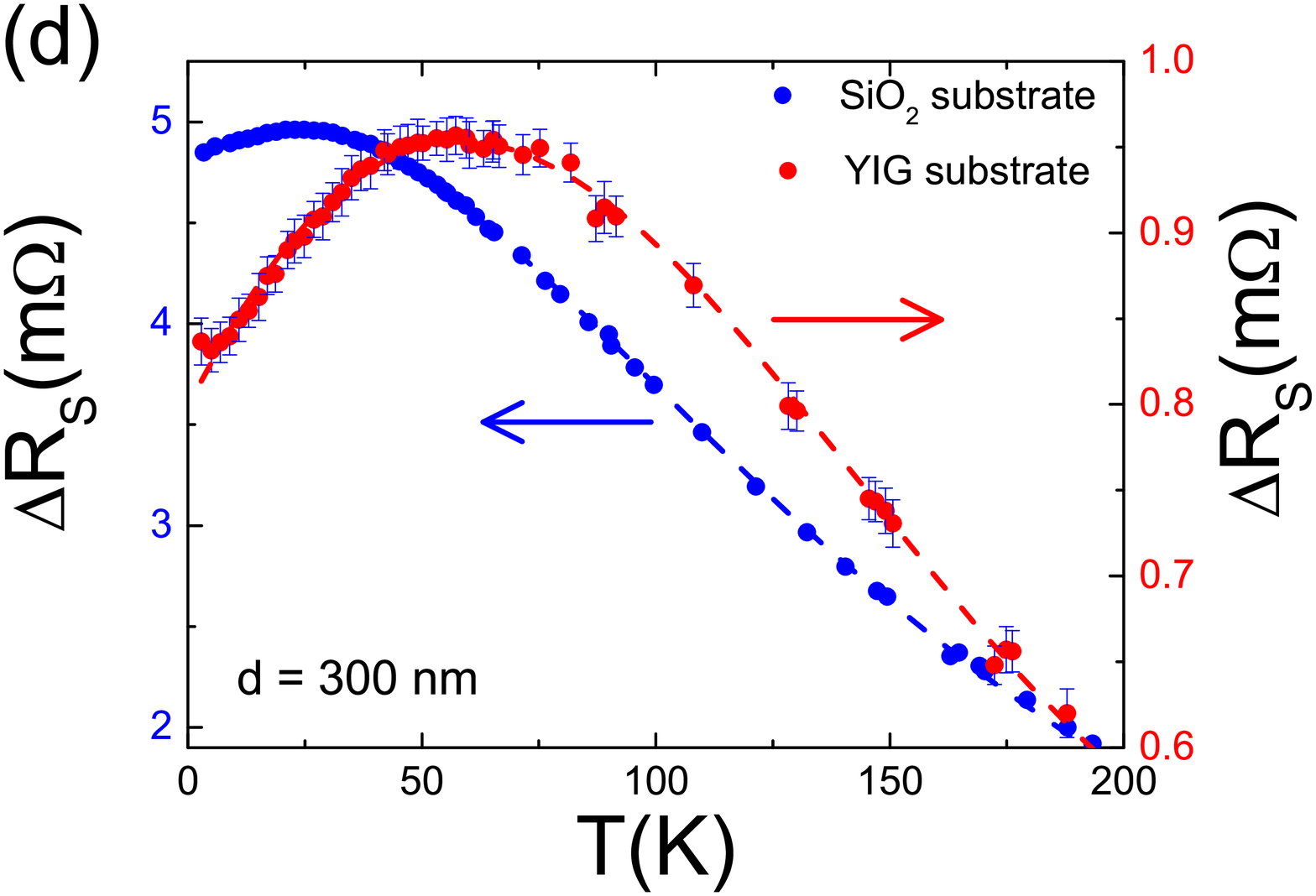}\\

\end{tabular}
\caption{(a) Nonlocal resistance $R_{NL}$  measured at 5 K for
LSVs fabricated on SiO$_2$ substrate with 3 nm MgO (blue) and 5 nm
EuS (red) capping. (b) Temperature-dependent spin signal $\Delta
R_{S}$ (symbols) for the same LSVs with injector-detector distance
$d$ = 500 nm. The dotted lines are guides to the eye. For all
temperatures, the device with magnetic EuS capping shows a smaller
value of the spin signal and the maximum of the spin signal is
shifted to higher temperatures. (c) Nonlocal resistance $R_{NL}$
measured at 10 K for LSVs fabricated on SiO$_2$ (blue) and YIG
(red) substrate. Both the LSVs were capped with AlO$_x$. (d)
Temperature dependent spin signal $\Delta R_{S}$ for the same LSVs
shown in (c) with injector-detector distance $d$ = 300 nm. Here
symbols are experimental data and dotted lines are guides to the
eye. As in (b), the device with the magnetic YIG substrate shows a
smaller spin signal throughout the temperature regime, as well as
a shift of the maximum to higher temperatures. }
\end{figure}
\end{center}
\end{widetext}
\begin{widetext}

\begin{center}
\begin{figure}[H]
\begin{tabular}{ll}
  \centering
  \includegraphics[width= 7 cm]{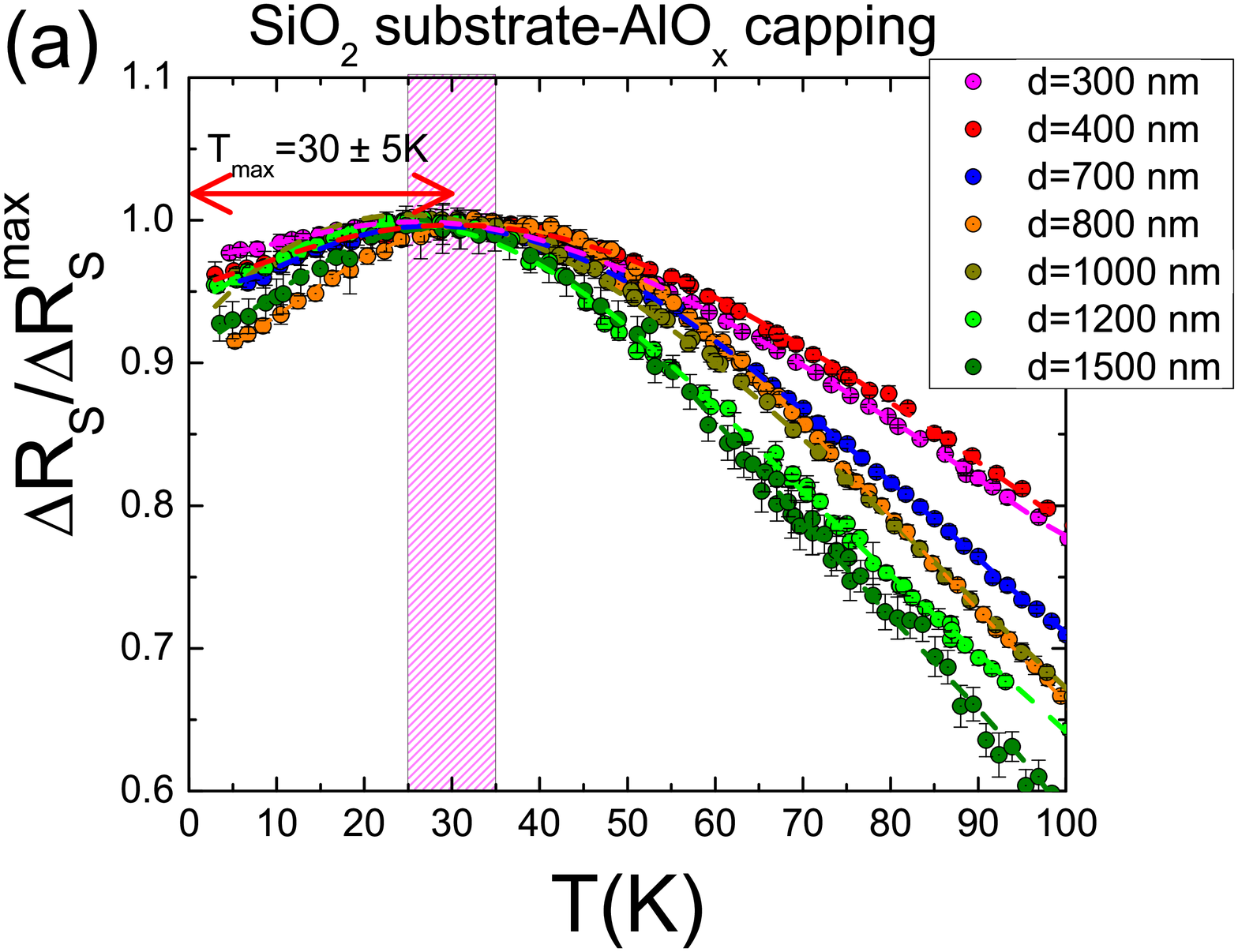}
&

  \includegraphics[width= 7 cm]{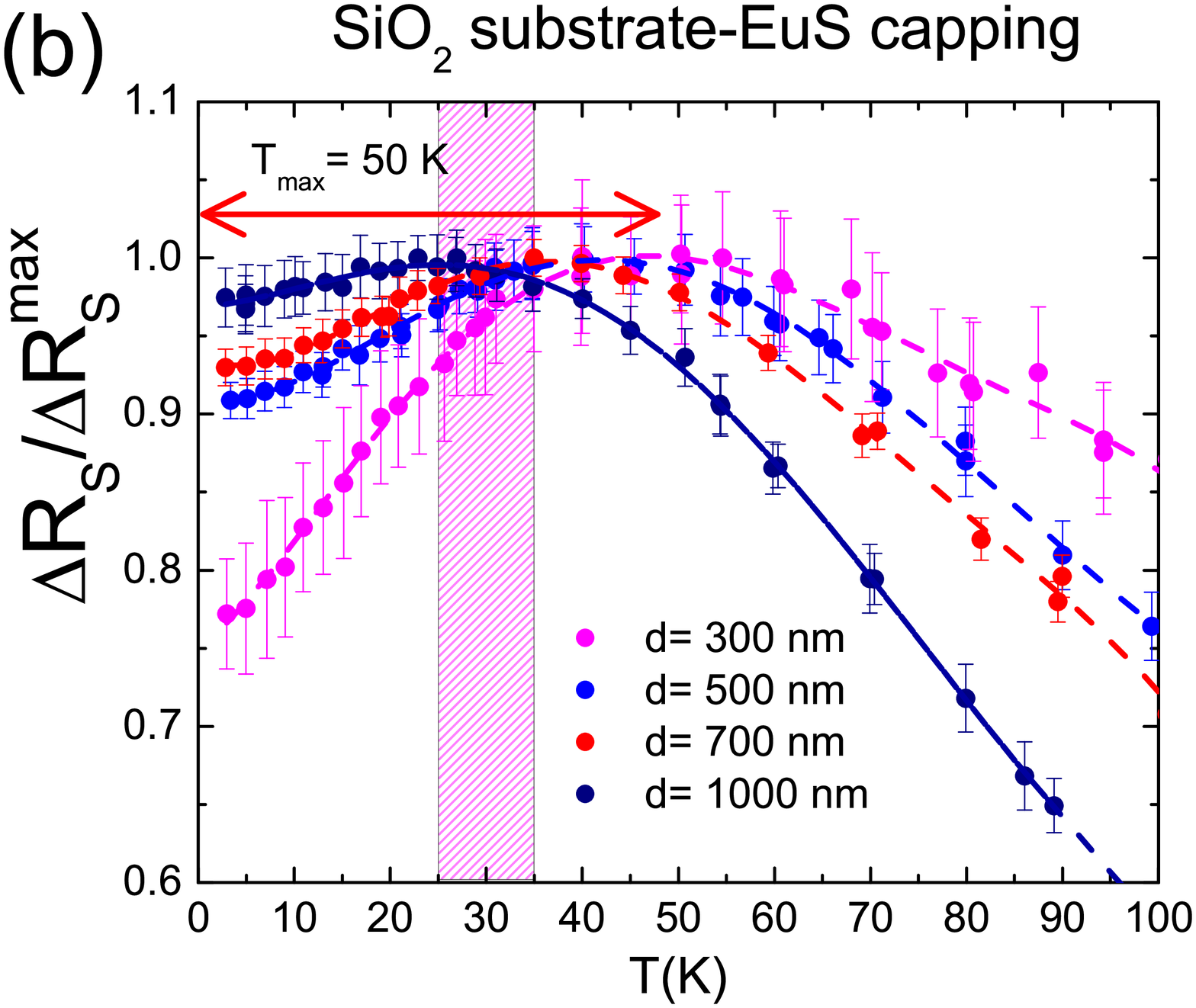}\\

 \centering
 \includegraphics[width= 7 cm]{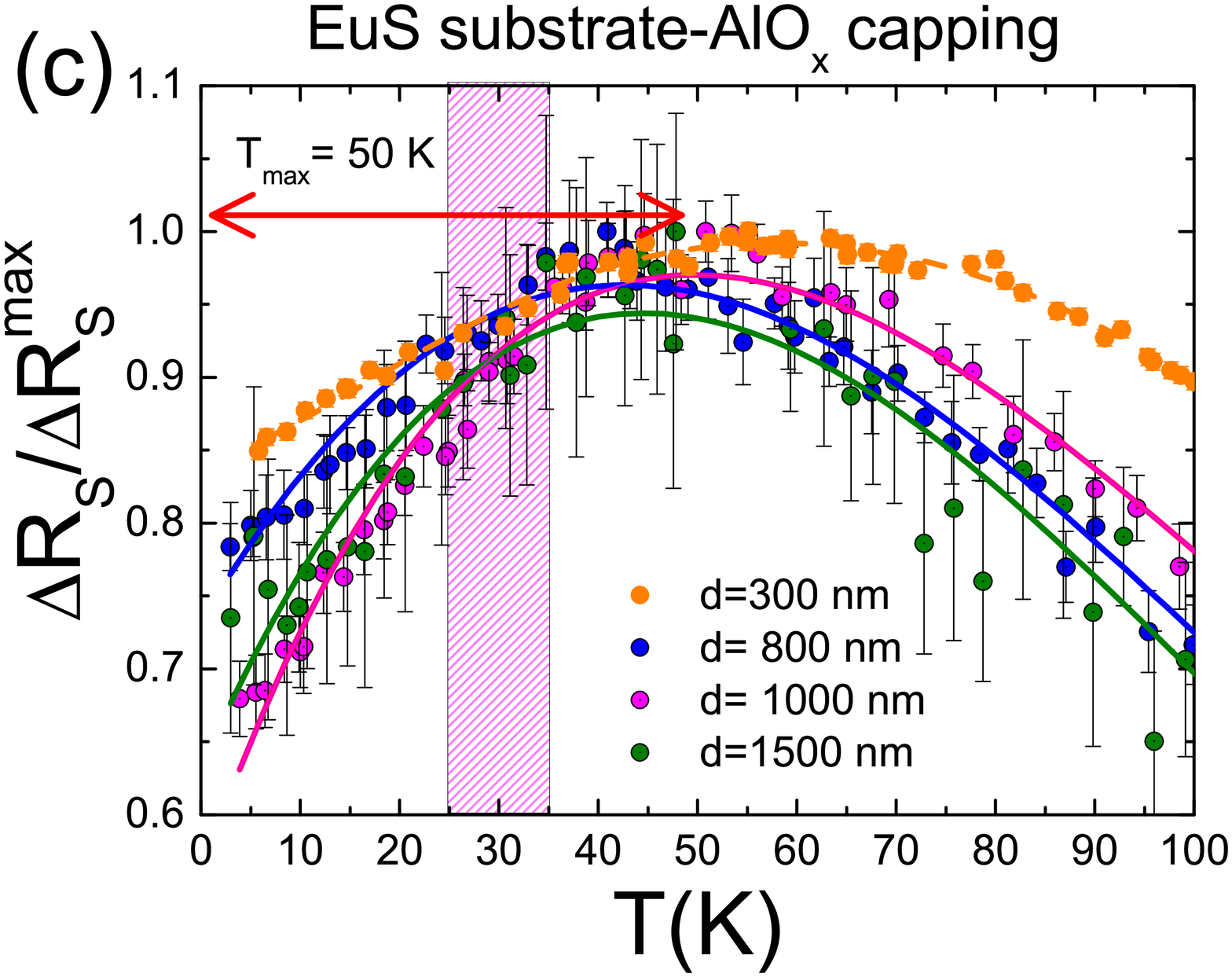}
&

 \includegraphics[width= 7 cm]{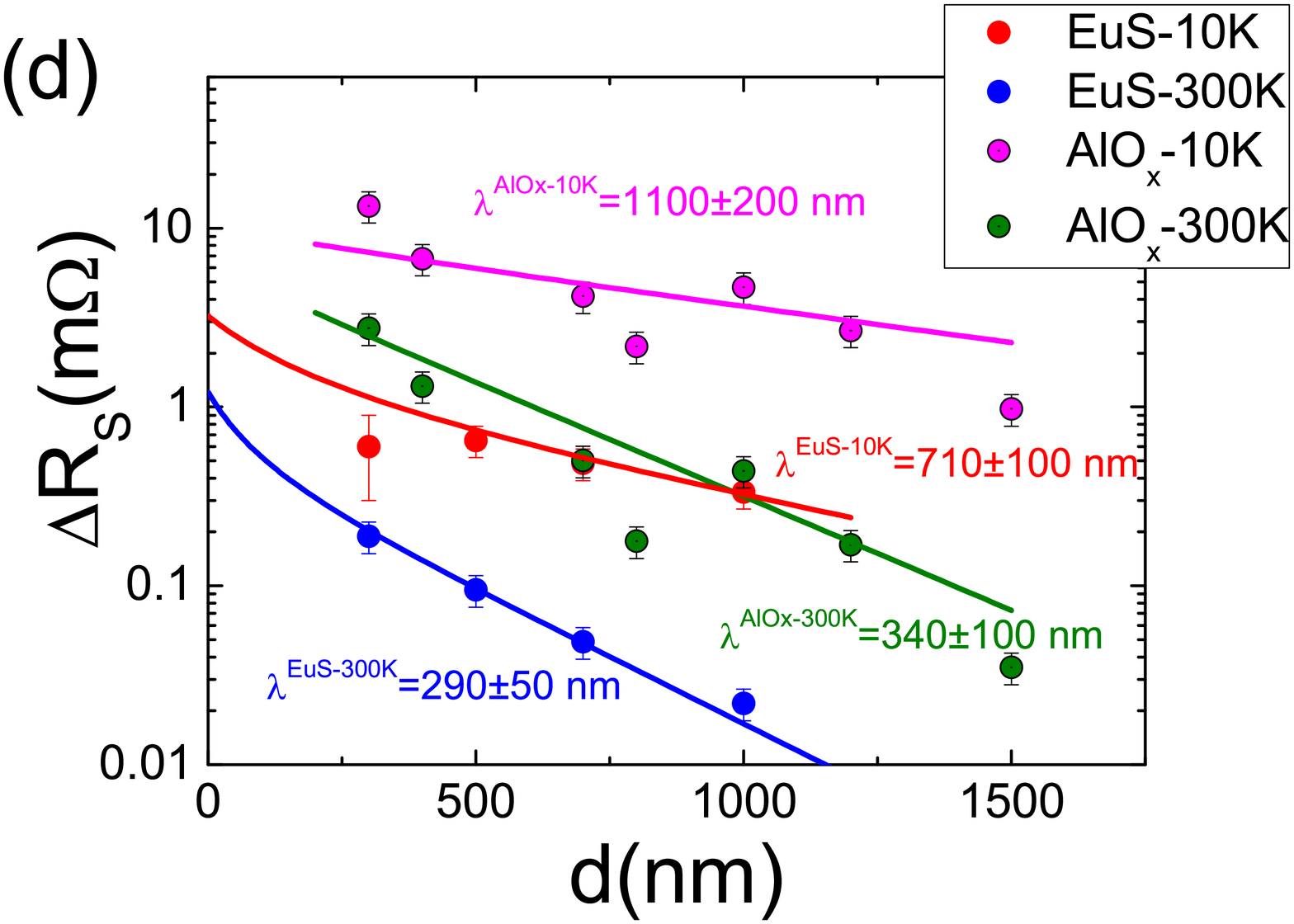}\\

\end{tabular}
\caption{Temperature dependence of spin signal in the Cu spin
transport channel for different injector-detector distance $d$ on
(a)SiO$_2$ substrate with AlO$_x$ capping, (b)SiO$_2$ substrate
with EuS capping and (c)EuS substrate with AlO$_x$ capping. All
the spin signals are normalized to their maximum value $\Delta
R_S^{max}$. The shadowed regions shows the expected temperature
ranges where the maximum spin signal is obtained for the reference
LSV devices. (d) Spin signal as a function of injector-detector
separation $d$ for the EuS and AlO$_x$ capped LSV devices. Solid
lines refer to the calculated dependence based on Eq. (2). }
\end{figure}
\end{center}
\end{widetext}

In order to create pure spin current in the spin transport channel
nonlocal configuration was used as shown in Fig. 1(b). When a
spin-polarized current is injected from the Py to Cu electrode
through a MgO tunnel barrier, a spin accumulation is built at the
interface between the Cu channel and the Py. This induced spin
accumulation at Py-MgO-Cu interface is expressed as, $\Delta \mu =
\mu ^ \uparrow - \mu ^ \downarrow$, with electrochemical
potentials $\mu ^ \uparrow$ and $\mu ^ \downarrow$ for up and down
spin electrons, respectively. This spin accumulation diffuses away
from the interface, creating a pure spin current in the Cu
nanowire which is detected as a voltage $V_{NL}$ by the second Py
electrode. The nonlocal resistance is defined from the
normalization of the detected voltage $V_{NL}$ to the injected
current $I$ as, $R_{NL} = V_{NL}/I$. The value of $R_{NL}$ changes
sign when the relative magnetization of the Py electrodes is
switched from parallel to antiparallel while sweeping applied
magnetic field along the long axis of Py. The change from positive
to negative $R_{NL}$ is defined as the \emph{spin signal}, $\Delta
R_S  = R_{NL}^{ \uparrow \uparrow }  - R_{NL}^{ \uparrow
\downarrow }$, which is proportional to the spin accumulation at
the detector. Therefore magnitude of the spin signal is a measure
of total spin relaxation time inside the spin transport channel.

Fig. 2(a) compares spin signal $\Delta R_{S}$ for two LSVs
fabricated on Si/SiO$_2$ substrate with EuS and MgO capping layer
and same electrode spacing $d$ =500 nm. Nonmagnetic MgO capping
layer could largely reduce the surface scattering providing ideal
reference devices for comparison with EuS capped
devices\cite{izuchi}. The nonlocal spin valve signal was measured
at 5 K below T$_{Curie}$ $\sim$ 15 K of EuS. The spin signal
$\sim$0.67 m$\Omega$ was observed for EuS capped devices which is
4.5 times less than spin signal $\sim$3.06 m$\Omega$ with MgO
capping.  A clear suppression of the nonlocal spin valve signal
can be observed. The amplitude of the spin signal also depends on
thickness of the tunnel barrier and therefore on the Cu/MgO/Py
interface resistance\cite{fukuma}. We found almost similar areal
interface resistance $R_{IA}^{EuS}$= 1.9 $f\Omega m^2$ and
$R_{IA}^{MgO}$= 0.6 $f\Omega m^2$ for the EuS and MgO capped
device at 10 K(Fig. 1(d)). These areal resistance values are
slightly higher than $R_{IA}$ ($\sim$0.3 $f\Omega m^2$ ) for spin
valve devices with no MgO tunnel barrier and correspond to a
semi-transparent tunnel barrier\cite{fukuma,lwang}. Fig. 2(c)
shows similar nonlocal resistance measurements to the one shown in
Fig. 2(a) comparing spin valves fabricated on YIG and SiO$_2$
substrate. For both devices, the same AlO$_x$ capping and distance
between two Py electrodes $d$ = 300 nm were used. The spin signal
for the spin valve on YIG substrate was found to be $\sim$6 times
smaller compared to the reference device fabricated on a
non-magnetic SiO$_2$ substrate. Similar suppression of the spin
signal on YIG substrates has been observed by other research
groups as well\cite{Dejene}. Fig. 2(b) and (d) shows temperature
dependence of spin signal for the same non-local spin valves shown
in Fig. 2(a) and (c), respectively. A uniform suppression of the
spin signal can be observed at all the temperatures. Furthermore
all devices exhibit a clear non-monotonous temperature-dependent
behavior, with a peak of the spin signal at low temperatures. In
both the reference spin valves either with MgO or AlO$_x$ capping
the peak in the spin signal was observed around $T_P$ $\approx$22
K. This low-temperature peak shifts to higher temperatures if the
spin-transport channel is in contact with either EuS ($T^{EuS}_P$
= 40 K) or YIG ($T^{YIG}_P$ = 60 K). As both EuS and YIG are
insulators no charge current will flow through them. We believe
this peak shift originates from interfacial effects as will be
discussed later on.

We further investigated the peak shift of the spin signal by
comparative temperature dependent measurements on lateral spin
valves with varying injector - detector distance $d$. Fig. 3(a)
shows temperature dependence of spin signal normalized to their
maximum value for several lateral spin valves on Si/SiO$_2$
substrate and AlO$_x$ capping. Irrespective of injector to
detector distance low temperature peak can be seen to appear at T
=30 $\pm$ 5 K. Also spin signal drops only $\sim$ 10$\%$ from the
maximum value at low temperature. The relationship between the
injector-detector distance $d$ and the spin signal $\Delta R_S$
can be described by a scaling law\cite{supp};
\begin{equation}
\Delta R_S [\lambda _{Cu} (T),d + \delta d] \approx K\Delta R_S
[\lambda _{Cu} (T),d]
\end{equation}
where $d$ is the injector-detector spacing and  $K$ is a
multiplication factor. This scaling law suggest that irrespective
of the reduction of the magnitude of spin signal $\triangle R$
with increased electrode spacing $d$ by $\delta d$, temperature
dependence of spin signal for all the spacing can be superimposed
onto each other with a constant multiplication factor $K$. Fig.
3(b) shows temperature-dependent spin signal measurements on EuS
capped LSV devices with varying injector-detector distance $d$.
Here a wide dispersion of the temperatures corresponding to the
peak maximum can be observed. In particular, for the LSV device
with  $d$ =300 nm the peak position is as high as 50 K. Besides
relative drop of the spin signal is quite high $\sim$ 25$\%$ from
its maximum value. It is obvious, that the curves in Fig. 3(b)
cannot be superimposed using Eq. (1), and the deviation from this
scaling suggest that additional temperature-dependent spin sinking
effects are present in the EuS capped LSV devices.

To further understand the anomalous low temperature behavior in
the nonlocal spin valves in contact with EuS, we investigated
another set of LSVs fabricated on Si/SiO$_2$/EuS(10 nm) thin film
and capped with AlO$_x$. In this set of samples, the
spin-transport channel might be in more uniform contact with the
smooth EuS film, as compared to the EuS-capped devices, which
might have non-uniform coverage. Fig. 3(c) shows normalized
temperature dependent spin signal of devices fabricated on EuS
substrate for different $d$. The peak positions are found at
temperatures T $\approx$ 50  K and a huge drop of the spin signal
up to $\sim$ 30$\%$ compared to maximum value can be seen at low
temperatures. In comparison with EuS capped LSV devices, the
devices grown on top of a EuS substrate show no significant $d$
dependence in the low temperature regime. A possible explanation
for the observed peak shifts to higher temperatures might
originate from effects like spin back flow into Py and anisotropic
spin absorption at the Py/Cu interfaces particularly when injector
- detector distance $d$ is comparable or shorter than spin
diffusion length $\lambda_s$ of
Cu\cite{Obrien-prb,izuchi-prb,izuchi-rev}. We believe the
anomalous low temperature behavior observed in our case does not
originate from these effects, as the shift would show a pronounced
dependence of the injector-detector distance $d$.

The variation of the spin signal amplitudes for devices having
different injector-detector gaps $d$ summarized for EuS and
AlO$_x$-capped devices is shown in Fig. 3(d). Based on the
one-dimensional spin diffusion model spin signal for a in-plane
magnetic field can be written as\cite{Takahashi};
\begin{equation}
\Delta R_S = 4R_{SCu} \frac{{\left[ {P_{MgO} \frac{{R_{SMgO}
}}{{R_{SCu} }} + P_{Py} \frac{{R_{SPy} }}{{R_{SCu} }}} \right]^2
{\mathop{\rm e}\nolimits} ^{ - d/\lambda _{Cu} } }}{{\left[ {1 +
2\frac{{R_{SMgO} }}{{R_{SCu} }} + 2\frac{{R_{SPy} }}{{R_{SCu} }}}
\right]^2  - {\mathop{\rm e}\nolimits} ^{ - 2d/\lambda _{Cu} } }},
\end{equation}
where $R_{SCu}  = \frac{{\rho _{Cu} \lambda _{Cu} }}{{t_{Cu}
w_{Cu} }}$, $R_{SPy}  = \frac{{\rho _{Py} \lambda _{Py} }}{{w_{Py}
w_{Cu} (1 - P_{Py}^2 )}}$, and $ R_{SMgO}  = \frac{{R_{IA}^{}
}}{{w_{Py} w_{Cu} (1 - P_{MgO}^2 )}}$ are the spin resistances of
Cu, Py and MgO interface, respectively. Here $\rho _i$ is the
resistivity, $P_i$ is the spin polarization, ${\lambda _i }$  is
spin diffusion length, $t_i$ is thickness, and $w_i$  is the width
of wire ($i$= Cu, Py or MgO). Here $R_{IA}$ stands for Py/MgO/Cu
areal interface resistance which was measured separately for each
devices and an average $R^{EuS}_{IA}$= 1.9 $f\Omega m^2$ and
$R^{AlO_x}_{IA}$= 27.9 $f\Omega m^2$ was used for fitting
purposes. Furthermore, the following values for Py at 10 K (300 K)
were assumed\cite{Sagasta}: $\rho_{Py}$=32 $\mu$$\Omega$-cm(44
$\mu$$\Omega$-cm), $\lambda _{Py}$= 3 nm(2.3 nm), and $P _{Py}$ =
0.39 (0.31). Fitting the data points shown in Fig. 3(d) to Eq. (2)
we extracted spin diffusion length of Cu at 10 K (300 K) on
SiO$_2$ substrate with AlO$_x$ capping as $\lambda_s$ $\approx$
1100$\pm$200 nm (340 $\pm$100 nm). These values are in good
agreement with reported values for Cu/Py lateral spin
valves\cite{Jedema,villamor-nlsv,taro}. A set of device properties
obtained from experiments and fitting Eq. (2) is  summarized in
Table 1. From the spin diffusion length $\lambda_s$ the spin
relaxation time can be calculated via, $ \tau _{sf} = \lambda
_{s}^2 /D $, where $D$ is the diffusion constant. The diffusion
constant was calculated from Einstein relation, $D = 1/N(E_F )e^2
\rho_{Cu}$ with $\rho_{Cu}$ the resistivity and $N(E_F )$
(=1.8$\times 10^{28}$ states/eVm$^3$\cite{Jedema}) is the density
of state at Fermi energy of Cu. The momentum relaxation time $\tau
_e$  due to defects was calculated using $\tau _e = 3/v_f^2 N(E_F
)e^2 \rho_{Cu}$. Here $v_f$ (= $1.57\times 10^6$ m/s) is the Fermi
velocity of Cu\cite{Jedema}.
\begin{widetext}

\begin{table}[!htbp]
\begin{tabular}{llllllllllllllllll}
\toprule

LSV &&$\rho_{Cu}$($\mu\Omega$-cm)&&$\lambda_s$(nm) &&$\tau_e$(fs)&&$\tau_{sf}$(ps)&&$\epsilon_{imp}$($\times 10^{-3}$)&& $p$($\times 10^{-4}$)\\
\colrule
AlO$_x$ capping &&2.06&&1100$\pm$200&&20.5&&71.8&&0.285&& 2.85 \\
EuS capping &&1.91&&710$\pm$100&&22.1&&27.7&&0.799&& 7.98 \\

\botrule \
\end{tabular}
\caption{Comparison of device parameters found from experiments
and fitting Eq. (2) to Fig. 3(d) for LSV devices with EuS and
AlO$_x$ capping: resistivity $\rho_{Cu}$, spin diffusion length
$\lambda_s$, momentum relaxation time $\tau_{e}$, spin relaxation
time $\tau_{sf}$, interfacial spin-flip parameter $\epsilon_{imp}$
and spin-flip probability $p$ measured at 10 K. } \label{tab1}
\end{table}

\end{widetext}

The spin-flip probability at the surface can be defined as, $p = 1
- e^{ - \epsilon_{imp}}$, where $\epsilon_{imp}$ =
$\tau_e$/$\tau_{sf}$ denotes the interfacial spin-flip
parameter\cite{bass}. From the values given in Table. 1, it is
obvious that EuS capped spin valves have higher spin flip
probability compared to AlO$_x$ capping. We believe this enhanced
spin-flip probability in the EuS capped LSV devices originates
from spin-flip scattering caused by the spin-orbit and exchange
fields at the Cu-EuS interface. Due to due to high atomic number
of Eu ($Z_{Eu}$ = 63) the spin-orbit field at the Cu-EuS interface
is larger than at Cu-AlO$_x$ interface ($Z_{Al}$ = 13). The
suppression of spin signal in EuS capped devices as compared to
the reference devices in full temperature range therefore point to
the spin-orbit field as the major cause of spin signal
suppression. However, anomalous scaling of the spin signal with
the injector-detector distance in the EuS capped LSV devices
cannot be explained with only spin-orbit fields. Therefore, in
order to explain total drop in the spin signal one must also
consider spin relaxation caused by interfacial exchange field and
spin sinking due to thermal magnons. Thus the spin diffusion
length in these LSV devices can be calculated considering
spin-mixing effects at the Cu-EuS interface. Following the
approach developed by Dejene \textit{et. al.}\cite{Dejene}
effective spin-relaxation length ($\lambda _{EuS}$) of Cu in the
EuS capped devices can be written as\cite{supp},

\begin{equation}
\frac{1}{{\lambda _{EuS}^2 }} = \frac{1}{{\lambda _{SiO2}^2 }} +
\frac{1}{{\lambda _{Sink}^2 }},
\end{equation}

where $\lambda _{SiO2}$ is the spin-diffusion length of Cu on
SiO$_2$ substrate and $\lambda _{Sink}$ denotes a length scale
which takes interfacial spin sinking effects into account. From
the value of  $\lambda _{Sink}$  the effective spin-mixing
conductance per area $G_s$ can be determined using the equation $
\frac{1}{{\lambda _{Sink}^2 }} = \frac{{2\rho _{Cu} G_s }}{{t_{Cu}
}}$, with Cu nanowire thickness $t_{Cu}$ = 100 nm and the Cu
resistivity $\rho _{Cu}$ = 1.91 $\mu\Omega$-cm. We found an
effective spin mixing conductance $G_s$ $\approx$3.03 $\times$
10$^{12}$ $\Omega^{-1} m^{-2}$ for the EuS capped LSV, $G_s$
$\approx$1.2 $\times$ 10$^{13}$ $\Omega^{-1} m^{-2}$ for the LSV
on EuS substrate, and $G_s$ $\approx$2.67 $\times$ 10$^{13}$
$\Omega^{-1} m^{-2}$ for LSV on YIG substrates. These values are
comparable to spin-mixing conductance $G_s$ $\approx$10$^{13}$
$\Omega^{-1} m^{-2}$ found for Al/YIG interfaces\cite{Dejene}.
Therefore, using interface engineering, it might be possible to
improve values of $G_s$ for Cu-EuS interface leading to an
effective spin current gating for spintronic devices.

\section{Discussion}

One of the main findings of our work is the observation of a
low-temperature peak in the spin signal of all fabricated
non-local spin valves, and furthermore the subsequent shift of the
peak to higher temperatures for devices in contact with a magnetic
insulator substrate or capping layer. The low-temperature peak in
the spin signal of lateral spin valves fabricated from high-purity
Cu has been observed by many research groups, but its physical
origin is still under debate, as its existence is not described by
the EY-mechanism used to explain spin relaxation in lateral spin
valve devices \cite{Elliott,Yafet}. Within the framework of the
EY-mechanism, the spin diffusion length is predicted to increase
monotonically with decreasing temperature. This prediction stands
in contrast to experimental observations where the spin signal of
lateral spin valves is usually found to decrease below 30 K, even
though the Cu resistivity, and thus the momentum relaxation time,
remain constant at low temperature (as demonstrated in Fig.~1(c)).
To explain the peak in the spin signal around $T \approx$30 K,
different mechanism including surface scattering
\cite{Erekhinsky,Kimura,villamor-nlsv,mihajlovic}, magnetic
impurities in bulk Cu or the vicinity of the Py/Cu
interface\cite{zhou,obrien,batley,watts,hamaya,Villamor-prb-nlsv,kim},
or changes in spin polarization of Cu/Py
interface\cite{Obrien-natcomm} have been proposed. The existence
of a small fraction (parts per million) of magnetic impurities in
bulk Cu is believed to cause $s-d$ spin-flip scattering associated
with the Kondo effect, which leads to a suppression of the
spin-diffusion length at low temperatures. This effect has been
extensively investigated using different magnetic impurities
inside a host nonmagnetic metal\cite{hamaya,obrien,batley}. One of
the observations from these experiments is that peak in the spin
signal appears at much higher temperature than the Kondo
temperature ($T_K$)\cite{Obrien-natcomm,batley,hamaya}.
Alternatively, surface spin-flip scattering has been shown to
contribute to an anomalous drop of the spin signal at low
temperatures\cite{Kimura,villamor-nlsv}. Interestingly,
experimental results indicate that the surface spin-flip rate
increases with the atomic number $Z$ of the atomic species at the
surface, suggesting that the spin-orbit field at the surface plays
a crucial role in the spin relaxation\cite{karube}. In the
pioneering work of Fert \textit{et. al.}\cite{fert}, it was
demonstrated that the spin relaxation due to spin-orbit scattering
is temperature-independent, in contrast to the
temperature-dependent exchange scattering. The fingerprints of
spin relaxation due to the interfacial exchange field can be seen
only at low temperature, where other temperature-dependent spin
relaxation processes are less prominent. Therefore,
temperature-dependent spin transport measurements on LSV devices
provide an unique way to extract information about the intrinsic
interfacial exchange field present at the FI/NM interface.

Recently, spin transport behavior of LSV in contact with a
ferromagnetic insulator has been theoretically
modeled\cite{Villamor-yig,Dejene}. When LSV is in proximity with a
ferromagnetic insulator spin angular momentum transfer occurs at
the interface and one has to consider additional boundary
condition for spin current density, which can be expressed
as\cite{Dejene,supp,Matthias}

\begin{equation}
\left. {j_s (\hat m)} \right|_{Interface}  = G_r \hat m \times
(\hat m \times \vec \mu _s ) + G_i (\hat m \times \vec \mu _s
)+G_s \vec \mu _s.
\end{equation}

Here  $\hat m = (m_x ,m_y ,0)$ is the unit vector parallel to
in-plane magnetization of EuS or YIG, $G_r$ and $G_i$ are real and
imaginary part of spin-mixing conductance per unit area, and $G_s$
is additional spin sinking term independent of magnetization
direction. According to this boundary condition when the spin
polarization of spin current in the metallic channel is
perpendicular to the magnetization, spin angular momentum can be
transferred to the magnetization of FI through spin transfer
torques. On the contrary when the polarization of the spin current
is aligned along the magnetization, the spin current cannot
penetrate into the insulator and is reflected back into the NM.
Successful observation of spin current modulation in LSVs on YIG
substrate has been attributed to a larger  value of the real part
of spin mixing conductance $G_r$ at the NM/YIG
interface\cite{Villamor-yig,Dejene}. However, in these experiments
role of interfacial exchange field were not addressed. One can
notice that in our measurements magnetization of EuS and both
injector-detector Py are all collinear, therefore, observed spin
memory loss cannot be explained by spin transfer torque. In
collinear case only the effective spin sink conductance $G_s$
becomes relevant. Temperature dependence of spin signal can be
explained considering temperature dependence of
$G_s$\cite{Matthias}. However, the effective spin sink conductance
is not clearly defined near and above the Curie
temperature\cite{Cornelissen}. Therefore, model based on Eq. (7)
may not be sufficient to explain temperature dependent spin signal
of LSVs in contact with EuS.

Recently, anomalous temperature dependent spin signal was also
observed in graphene spin valves in contact with
YIG\cite{Simranjeet}. Alternative approach was developed
considering randomly fluctuating exchange fields. In our
experiments anomalous low temperature behavior most likely appears
due to spin relaxation caused by interfacial exchange field at the
Cu-EuS (or Cu-YIG) interface. At finite temperatures, the Eu
moments (Fe-moment in case of YIG) fluctuate around the local
equilibrium state producing fluctuating exchange field ($B_{ex}$)
at the interface. Although the Curie temperature of EuS is
$\sim$15 K thermal fluctuation of Eu moments can survive at
temperatures above $T_C$ as seen in some ferromagnets\cite{Qin}.
These fluctuating exchange field exponentially decay with time as,
$ \Delta \left\langle {\vec B_{ex} (t).\vec B_{ex}(t - t')}
\right\rangle _t \propto \exp ( - t/\tau _c )$, where $\tau _c$ is
the fluctuation correlation time. According to Matthiessen's rule
total spin-flip time after considering contribution from
fluctating exchange field can be written as,

\begin{equation}
\frac{1}{{\tau _{sf} (T)}} = \frac{1}{{\tau _{sf}^i (0)}} +
\frac{1}{{\tau _{sf}^p (T)}} + \frac{1}{{\tau _{sf}^{ex} (T)}}.
\end{equation}
Here $\tau _{sf}^i $, $\tau _{sf}^p $ and $\tau _{sf}^{ex}$
represents spin-flip scattering time due to non-magnetic
impurities or defects, phonons and interfacial exchange field,
respectively. According to Elliott-Yafet theory spin relaxation
($\tau _{sf}^i$) from nonmagnetic defects  are temperature
independent\cite{Elliott,Yafet}. Also at low temperature phonon
contribution to the spin relaxation rate vanishes. This leaves the
fluctuating exchange field as the only temperature-dependent
contribution to the spin relaxation. Based on the exchange-field
model developed by McCreary \emph{et.\ al.}\cite{McCreary}, the
spin-relaxation rate due to the fluctuating exchange field can be
written as,

\begin{equation}
\frac{1}{{\tau _{sf}^{ex} (T)}} = \frac{{[\Delta B_{ex}(T)]^2
}}{{\tau _c(T) }}\frac{1}{{[B_{ap}  + B_{ex}(T) ]^2  + \left[
{\frac{\hbar }{{g\mu _B \tau _c(T) }}} \right]^2}},
\end{equation}

where $\Delta B_{ex}$ is the rms value of exchange field
fluctuation amplitude in the transverse direction, $B_{ap} $ is
applied magnetic field, $B_{ex} $ is the magnitude of exchange
field and $\tau _c $ is the fluctuation correlation time. The
spin-relaxation rate in Eq. [6] is temperature dependent as
interfacial exchange field which is determined by magnetization is
temperature dependent. Recent measurements of our group on
metallic spin glass like CuMnBi further showed that the spin Hall
angle decreases at temperatures $T^*$ well above the materials
spin-glass temperature $T_g$\cite{nimmi-jpsj}. This unusual
temperature dependence has been understood considering magnetic
fluctuations of Mn moment which can be detected more sensitively
by spin Hall effect\cite{ziman}. All these experiment suggest
temperature dependent measurement involving spin current can be
used to study fluctuating interfacial exchange field in a more
precise manner than charge transport methods.

\section{Conclusions}
In summary, we presented a detailed study of temperature-dependent
spin signal of non-local spin valves in contact with magnetic EuS
or YIG, and compared these to reference devices adjacent to
nonmagnetic MgO or AlO$_x$. We found that the spin signal is
suppressed for devices in contact with a magnetic EuS or YIG
layer, and attribute this suppressed signal to an enhanced surface
scattering originating from the spin-orbit fields at the interface
to the magnetic capping layer or substrate. Besides spin signal
suppression we found widely observed low temperature peak in the
spin signal was shifted to much higher temperature compared to
control LSV devices. Furthermore, the spin signal was found to
scale in an anomalous way with the injector-detector distance of
these LSV devices. We believe these additional temperature
dependent spin sinking effects arise due to fluctuating exchange
field at the Cu-EuS or Cu-YIG interfaces. The strength of this
exchange field can be determined in the future using Hanle
spin-precession experiments. Moreover anomalous temperature
dependence observed in our spin valve devices provide an
unambiguous evidence for the presence of exchange field at the
EuS-Cu interface. With careful engineering of the interface
quality it might be possible to demonstrate magnetically
controlled modulation of the spin current in these devices. The
ongoing development and miniaturization in the field of
spintronics demands ultrathin magnetic films which are stable
under ambient conditions and easy integrability into nanoscale
metallic devices. Our experimental results suggest ferromagnetic
insulator EuS can be a handy material for nanoscale spintronics
devices.

\noindent \textbf{Acknowledgments}\\
This work was supported by a Grant-in-Aid for Scientific Research
on Innovative Area, "Nano Spin Conversion Science" (Grant No.
26103002). SPD thanks Swedish Research Council project grant (No.
2016-03658). We would like to thank Dr T. Nakamura and Prof S.
Katsumoto for the use of the lithography facilities. We also thank
Dr Na\"{e}mi Riccarda Leo for proofreading the manuscript.

\newpage


\begin{thebibliography}{100}
\bibitem{otani}
Y. Otani, M. Shiraishi, A. Oiwa, E. Saitoh and S. Murakami, Nature
Physics \textbf{13}, 829 (2017).

\bibitem{Soumyanarayanan}
A. Soumyanarayanan, N. Reyren, A. Fert, C. Panagopoulos, Nature
\textbf{539}, 509 (2016).

\bibitem{Hellman}
F. Hellman, A. Hoffmann, Y. Tserkovnyak, G. S. Beach, E. E.
Fullerton, C. Leighton, A. H. MacDonald, D. C. Ralph, D. A. Arena,
H. A. Dürr, P. Fischer, J. Grollier, J. P. Heremans, T. Jungwirth,
A. V. Kimel, B. Koopmans, I. N. Krivorotov, S. J. May, A. K.
Petford-Long, J. M. Rondinelli, N. Samarth, I. K. Schuller, A. N.
Slavin, M. D. Stiles, O. Tchernyshyov, A. Thiaville, and B. L.
Zink, Rev. Mod. Phys. \textbf{89}, 025006 (2017).

\bibitem{edelstein}
V. M. Edelstein, Solid State Commun. \textbf{73}, 233 (1990).

\bibitem{rojasan}
J. C. Rojas Sánchez, L. Vila, G. Desfonds, S. Gambarelli, J. P.
Attané, J. M. De Teresa, C. Magén and A. Fert, Nature
Communications \textbf{4}, 2944 (2013).

\bibitem{Tserkovnyak}
Y. Tserkovnyak, A. Brataas, G. E. W. Bauer, and B. I. Halperin,
Rev. Mod. Phys. \textbf{77}, 1375 (2005).

\bibitem{Ralph}
D. C. Ralph and M. D. Stiles, J. Magn. Magn. Mater. \textbf{320},
1190 (2008).

\bibitem{Uchida}
K. Uchida, J. Xiao, H. Adachi, J. Ohe, S. Takahashi, J. Ieda, T.
Ota, Y. Kajiwara, H. Umezawa, H. Kawai, G. E. W. Bauer, S.
Maekawa, and E. Saitoh, Nat. Mater. \textbf{9}, 894 (2010).

\bibitem{Flipse}
J. Flipse, F. K. Dejene, D. Wagenaar, G. E. W. Bauer, J. Ben
Youssef, and B. J. van Wees, Phys. Rev. Lett. \textbf{113}, 027601
(2014).

\bibitem{Nakayama}
H. Nakayama, M. Althammer, Y.-T. Chen, K. Uchida, Y. Kajiwara, D.
Kikuchi, T. Ohtani, S. Geprägs, M. Opel, S. Takahashi, R. Gross,
G. E. W. Bauer, S. T. B. Goennenwein, and E. Saitoh, Physical
Review Letters \textbf{110}, 206601 (2013).



\bibitem{Goennenwein}
S. T. B. Goennenwein, R. Schlitz, M. Pernpeintner, K. Ganzhorn, M.
Althammer, R. Gross, and H. Huebl, Appl. Phys. Lett. \textbf{107},
172405 (2015).

\bibitem{Elliott}
R. J. Elliott, Phys. Rev. \textbf{96}, 266 (1954).

\bibitem{Yafet}
Y. Yafet, Solid State Phys. \textbf{14}, 1 (1963).

\bibitem{Villamor-yig}
E. Villamor, M. Isasa, S. Vélez, A. Bedoya-Pinto, P. Vavassori, L.
E. Hueso, F. S. Bergeret, and F. Casanova, Phys. Rev. B
\textbf{91}, 020403 (2015).

\bibitem{Dejene}
F. K. Dejene, N. Vlietstra, D. Luc, X. Waintal, J. Ben Youssef,
and B. J. van Wees, Phys. Rev. B \textbf{91}, 100404(R)(2015).

\bibitem{Maekawa-rev} S. Maekawa, H. A. Adachi, K. Uchida, J.
Ieda, and E. Saitoh, J. Phys. Soc. Jpn. \textbf{82}, 102002
(2013).


\bibitem{Miao-rev}
G.X. Miao, J.S. Moodera, Phys. Chem. Chem. Phys. \textbf{17}, 751
(2015).

\bibitem{Strambini}
E. Strambini, V. N. Golovach, G. De Simoni, J. S. Moodera, F. S.
Bergeret, and F. Giazotto,Phys. Rev. Materials \textbf{1}, 054402
(2017).

\bibitem{moodera}
J. S. Moodera, T. S. Santos, and T. Nagahama, J. Phys.: Condens.
Matter \textbf{19}, 165202 (2007).

\bibitem{hao}
X.Hao, J.S.Moodera, and R. Meservey, Phys. Rev. B \textbf{42},8235
(1990).

\bibitem{pwei}
P. Wei, F. Katmis, B. A. Assaf, H. Steinberg, P. Jarillo-Herrero,
D. Heiman, and J. S. Moodera, Phys. Rev. Lett. \textbf{110},
186807 (2013).

\bibitem{wei}
P. Wei, S. Lee, F. Lemaitre, L. Pinel, D. Cutaia, W. Cha, F.
Katmis, Y. Zhu, D. Heiman, J. Hone, J. S. Moodera, and C.-T. Chen,
Nat. Mater. \textbf{15}, 711 (2016).

\bibitem{zhao}
C. Zhao, T. Norden, P. Zhang, P. Zhao, Y. Cheng, F. Sun, J. P.
Parry, P. Taheri, J. Wang, Y. Yang, T. Scrace, K. Kang, S. Yang,
G.-x. Miao, R. Sabirianov, G. Kioseoglou, W. Huang, A. Petrou, and
H. Zeng, Nat. Nanotechnol. \textbf{12}, 757 (2017).

\bibitem{Katmis}
F. Katmis, V. Lauter, F. S. Nogueira, B. A. Assaf, M. E. Jamer, P.
Wei, B. Satpati, J. W. Freeland, I. Eremin, D. Heiman, P.
Jarillo-Herrero, and J. S. Moodera, Nature (London) \textbf{533},
513 (2016).

\bibitem{Leutenantsmeyer}
C. Leutenantsmeyer, A. A. Kaverzin, M. Wojtaszek, and B. J. van
Wees, 2D Materials \textbf{4}, 014001 (2017).


\bibitem{zinn}
W. Zinn, J. Magn. Magn. Mater.\textbf{3}, 23(1976).


\bibitem{Idzuchi-EuS}
H. Idzuchi, Y. Fukuma, H. S. Park, T. Matsuda, T. Tanigaki, S.
Aizawa, M. Shirai, D. Shindo, and Y. Otani, Appl. Phys. Exp.
\textbf{7}, 113002 (2014).

\bibitem{wolf}
M. J. Wolf, C. Surgers, G. Fischer, T. Scherer, D. Beckmann,
J.Magn. Magn. Mater. \textbf{368},499 (2014).


\bibitem{supp}
See Supplemental Material at http://link.aps.org/ supplemental/
for LSV with no MgO tunnel barrier, Energy-dispersive X-ray(EDX)
spectra to validate presence of EuS on the EuS capped device, Note
on scaling Eq. [1], and Calculation of effective spin relaxation
length of Cu in LSV in contact with FI (EuS or YIG).




\bibitem{izuchi}
H. Idzuchi, Y. Fukuma, L. Wang, and Y. Otani, Appl. Phys. Lett.
\textbf{101}, 022415 (2012).

\bibitem{fukuma}
Y. Fukuma, L. Wang, H. Idzuchi, S. Takahashi, S. Maekawa, and Y.
Otani, Nature Materials \textbf{10}, 527 (2011).

\bibitem{lwang}
L. Wang, Y. Fukuma, H. Idzuchi, and Y. Otani, J. Appl. Phys. 109,
07C506 (2011).

\bibitem{Obrien-prb}
L. O'Brien, D. Spivak, N. Krueger, T. A. Peterson, M. J. Erickson,
B. Bolon, C. C. Geppert, C. Leighton, and P. A. Crowell, Phys.
Rev. B \textbf{94}, 094431 (2016).





\bibitem{izuchi-prb}
H. Idzuchi, Y. Fukuma, S. Takahashi, S. Maekawa, and Y. Otani
Phys. Rev. B \textbf{89}, 081308(R) (2014).

\bibitem{izuchi-rev}
H. Idzuchi, Y. Fukuma, Y. Otani, Physica E \textbf{68}, 239
(2015).

\bibitem{Takahashi}
S. Takahashi and S. Maekawa, Sci. Technol. Adv. Mater.\textbf{9},
014105 (2008).

\bibitem{Sagasta}
E. Sagasta, Y. Omori, M. Isasa, Y. Otani, L. E. Hueso, and F.
 Casanova, Appl. Phys. Lett. {\bf 111}, 082407 (2017).

\bibitem{Jedema}
F. J. Jedema, M. S. Nijboer, A. T. Filip, and B. J. van Wees,
Phys. Rev. B \textbf{67}, 085319 (2003).

\bibitem{villamor-nlsv}
E. Villamor, M. Isasa, L. E. Hueso, and F. Casanova, Phys. Rev. B
\textbf{87}, 094417 (2013).

\bibitem{taro}
T. Wakamura, K. Ohnishi, Y. Niimi, and Y. Otani, Applied Physics
Express \textbf{4}, 063002(2011).

\bibitem{bass}
J. Bass and W. P. Pratt Jr., J. Phys. Condens. Matter \textbf{19},
183201 (2007).

 \bibitem{Erekhinsky}
 M. Erekhinsky, A. Sharoni, F. Casanova, and I. K. Schuller, Appl. Phys. Lett. \textbf{96}, 022513 (2010).


 \bibitem{Kimura}
 T. Kimura, T. Sato, and Y. Otani,Phys. Rev. Lett. \textbf{100}, 066602 (2008).


\bibitem{mihajlovic}
G. Mihajlovic, J. E. Pearson, S. D. Bader, and A. Hoffmann, Phys.
Rev. Lett. \textbf{104}, 237202 (2010).

\bibitem{zhou}
H. Zhou and Y. Ji, Appl. Phys. Lett. \textbf{101}, 082401 (2012).

\bibitem{obrien} L. \'{O}Brien, D. Spivak, J. S. Jeong, K. A. Mkhoyan,
P. A. Crowell, and C. Leighton, Phys. Rev. B \textbf{93}, 014413
(2016).

\bibitem{batley}
J. T. Batley, M. C. Rosamond, M. Ali, E. H. Linfield, G. Burnell,
and B. J. Hickey, Phys. Rev. B \textbf{92}, 220420(R) (2015).


\bibitem{watts}
J. D. Watts, J. S. Jeong, L. OfBrien, K. A. Mkhoyan, P. A.
Crowell, and C. Leighton, Appl. Phys. Lett. \textbf{110}, 222407
(2017).


\bibitem{hamaya}
K. Hamaya, T. Kurokawa, S. Oki, S. Yamada, T. Kanashima, and T.
Taniyama, Phys. Rev. B \textbf{94}, 140401(R) (2016).

\bibitem{Villamor-prb-nlsv}
E. Villamor, M. Isasa, L. E. Hueso, and F. Casanova, Phys. Rev. B
\textbf{88}, 184411 (2013).


\bibitem{kim}
K.-W. Kim, L. O'Brien, P. A. Crowell, C. Leighton, and M. D.
Stiles, Phys. Rev. B \textbf{95}, 104404 (2017).

\bibitem{Obrien-natcomm}
L. O'Brien, M. J. Erickson, D. Spivak, H. Ambaye, R. J. Goyette,
V. Lauter, P. A. Crowell, and C. Leighton, Nat. Commun.
\textbf{5}, 3927 (2014).

\bibitem{karube}
S. Karube, H. Idzuchi, K. Kondou, Y. Fukuma, and Y. Otani, Appl.
Phys. Lett. \textbf{107}, 122406 (2015).


\bibitem{fert}
A. Fert, J.-L. Duvail, and T. Valet, Phys. Rev. B \textbf{52},
6513 (1995).

\bibitem{Matthias} Matthias Althammer, arXiv:1802.08479

\bibitem{Cornelissen} L. J. Cornelissen, K. J. H. Peters, G. E.
W. Bauer, R. A. Duine, and B. J. van Wees, Phys. Rev. B 94, 014412
(2016).

\bibitem{Simranjeet}
S. Singh, J. Katoch, T. Zhu, K.-Y. Meng, T. Liu, J. T. Brangham,
F. Yang, M. E. Flatté, and R. K. Kawakami, Phys. Rev. Lett.
\textbf{118}, 187201 (2017).

\bibitem{Qin} H. J. Qin, Kh. Zakeri, A. Ernst, and J. Kirschner,
Phys. Rev. Lett. \textbf{118}, 127203 (2017).

\bibitem{McCreary}
K. M. McCreary, A. G. Swartz, W. Han, J. Fabian, and R. K.
Kawakami, Phys. Rev. Lett. 109, 186604 (2012).



\bibitem{nimmi-jpsj}
Y. Niimi, D. Wei, and Y. Otani, J. Phys. Soc. Jpn. \textbf{86},
011004 (2017).

\bibitem{ziman}
T. Ziman, B. Gu, and S. Maekawa, J. Phys. Soc. Jpn. \textbf{86},
011005 (2017).




\end{thebibliography}
\end{document}